\documentstyle[aps,epsf,prl,twocolumn]{revtex}
\newcommand{\om}{{\omega}}

\newcommand{\bel}{\begin{equation}\label}
\newcommand{\ee}{\end{equation}}

\begin{document}
\twocolumn[{
\draft

\title {
\Large {\bf Extended and localized vibration
models in  disordered systems}
}

\author { Barak Galanti and Zeev Olami }
\address {
Department of~~Chemical Physics,\\
The Weizmann Institute of Science,
Rehovot 76100, Israel
}
\maketitle
\widetext

\begin{abstract}
We discuss vibrational localization problems in glasses and disordered media
in this paper.  It is claimed that the essence of the localization 
problem is already observed in disordered lattice models.
These kinds of vibrations belong to a different universality class than 
bonded electrons.
Specifically, The eigenvectors are extended even 
in two dimensions. Moreover, the correlation exponent does not diverge in the 
transition from localized to delocalized states.
Furthermore, the volume of the extended states is scaled according to
the distance from the transition as $V\sim |\omega-\omega_c| L^d$.
Interestingly, boson peaks can be observed in the density of states 
in both two and three dimensions.  
We studied the eigenstates of this problem and
analyzed the scattering effects in these lattices. 
Importantly, we found that in two dimensions 
the boson peak, the localization edge, and the beginning
of anomalous scattering are at the same frequency.
In three dimensions, however,
there are three separate regions: (1) localized, (2) weakly scattered, and 
(3) anomalously scattered.
Finally, we discuss the relevance of this study 
to actual experiments and glasses. 

\noindent
  
%\end{titlepage}
\end{abstract}
\leftskip 54.8pt
\pacs{PACS numbers: 63.20 Localized Modes,
                    63.50 Vibrational States in Disordered Systems,
                    73.23 Theories and Modes for Localized States}

}]
\narrowtext

\section{Introduction}

It is well-known that a disorder in the properties of a physical system
can in turn induce localization and scattering effects.
For the past three decades, disordered electronic systems have been 
the main focus of much research, having received enormous attention as 
manifested by the numerous books and reviews on the subject[1-10]. 
In contrast, though the disordered vibration problem
is as important as the electronic one,
we have found almost no discussion on the analogous problem of 
phonon localization. Though the vibrational properties of disordered 
solids were reviewed in several papers \cite{vibold}, subjects like 
localization, scattering, and properties of the eigenstates were 
not analyzed. 
There are various reasons for this.
The main one is that localization effects for electrons 
are more pronounced physically for statistical reasons.
They occur at a well-defined 
Fermi energy state, so it is much easier to observe them 
under experimental conditions.
In contrast, the localization effects of phonons are often 
obscured behind the calculations and the statistics; therefore,
their role in influencing properties like heat
conductivity or scattering is much less clear. We believe therefore that the
theoretical understanding of these localization effects is crucial to the
understanding of glasses. 

Unfortunately, we found almost no systematic treatment of 
vibration in disordered media above one-dimension in the literature. 
However, there is some literature relating to two main issues.
The first is a  bonded network with cuts in the connections,
which is the percolation problem.
In such a problem anomalous fractal
excitations called fractons may appear instead of the usual 
long-range phonons.
This phenomenon was first mentioned in \cite{ale82} and has been repeatedly
observed in experiments and simulations.
The second type of problem comprises simulations of 
`real' glasses \cite{GlassGen}. 
Though such simulations do yield some information 
about the relevant systems, it is rather difficult to obtain any
concrete ideas because of various reasons. 
For example, simulations of glasses are completely uncontrolled in as 
much as the source of the disorder and its size are 
unknown, since the glass is defined by a dynamical relaxation algorithm.
It is indeed possible to study these glasses and probe the degree of 
elastic and structural disorder; however, the disorder is still 
uncontrollable. Clearly, the effects of the disorder are 
manifested in  many experimental measurements in glasses, but it is 
difficult to make a reasonable and rational interpretation of the data. 

Disorder in glasses and other materials can originate from 
various physical sources. Structural characteristics can change the topology 
of the material and thus affect
the vibrations. Another major source of disorder is elastic 
variability in the local elastic constants. In simple glasses this is the 
main source of disorder, which is caused by the quenched pressure that 
remains in the glass after the quench from 
the liquid. Such quenched pressures are correlated with  
the existence of second-order gradients
that generate large variations  in the 
local elastic constants \cite{ale98,KO99}.
Another important component for disorder is  the mass
disorder that occurs when two or more kinds of atoms
exist in the lattice. 

In glasses there is a set of anomalous thermodynamic properties 
that do not appear in  crystals \cite{Elliot,FA}.
Examples of such behavior are anomalous heat conductivity
and excess vibrations at low frequencies termed {\it boson peak}
in the literature\cite{Elliot,FA}.
Scattering mechanisms are thought to mainly contribute to the
depletion of the thermal transport in glasses.
A crossover in the scattering of the glass is expected to be observed on 
a characteristic scale which is usually identified with
the Ioffe-Regel limit,
where the scattering length is of the same order as the wave-length
\cite{GlassGen,Pa99}. In fact we observe two basic 
scales in three dimensions 
as will be discussed later. Both of them do not fit the  I-R demand
 but they are in the same range.  
In the second part of this  paper we focus on scattering
effects.
Since getting experimental \cite{Courtens}
and numerical information is relatively  
difficult \cite{GlassGen}; it is easier to use lattice 
models to obtain  information about the vibrational properties.
Indeed, we shall show here that many of the basic 
phenomena observed in glasses are also observed in lattice models.

In the one-dimensional case, it is  known that the vibrations
are localized \cite{Ea,Derr84,ITZ10}.  
However, the localization length in these systems is scaled as the 
inverse square of the frequency. So at low frequencies, we observe
phonon-like modes with a localization length much larger than
their wavelength, and with a low frequency spectral density 
of order one. 
In a few theoretical papers on this subject \cite{SJ83,CZ88},
it was contended that the situation at higher dimensions is analogous to that  
in electrons. We present here sufficient numerical evidence to 
show that this is not the case. 

The vibration problem generally consists of a vector translation
problem.
A simpler version of the problem is the problem of scalar vibrations where
there is a scalar translation at any point. This is a realistic problem
in one and two dimensions (e.g., a membrane). 
However, scalar models are, as we shall see, interesting enough 
theoretically that we should start with a discussion 
of such models.

There are in fact, there are three non-equivalent 
physical systems which, under certain constraints would
give the same dynamical matrix: 
(1) the vibration problem (phonons) in disordered
media, (2) free electrons with randomness in their transmission coefficients,
and (3) the random diffusion problem.
The dynamical structure of  these problems is similar when the 
following constraints are fulfilled: (1) the transmission coefficients 
are symmetric and positive, (2) the relevant dynamical variables are  
scalar, and (3) the  masses are constant.

Variants from these constraints, i.e. the vectorial case, the
negative connectivity, and the variable masses, display different results
and will be discussed in other papers \cite{OG99}.   
However, the time dependence, the statistics, and physical properties of
these systems are very different. 
If any of the constraints are broken, we can expect to observe a different
behavior, as will be described later in detail.

In this paper we extend our discussion of an earlier, shorter paper on the 
subject \cite{GO99}. In section II we describe the concepts of
localization in electronic theory, and in general.
In section III we demonstrate that the dynamical matrices of the three 
previously mentioned systems are the same, when the parameter 
constraints that we mentioned are satisfied.
In section IV we discuss the spectra and localization of
the scalar models. We first analyze the 
one-dimensional case and then proceed
to  analyze numerically lattice problems that are analogous
to the models studied in the electronic case[1-10]. 
By increasing the lattice size we can characterize the eigenstates. 
The participation ratios of the states are used to measure the degree of
localization.  A localization transition between extended
and localized states is observed both  in two and three dimensions.
Furthermore, in two dimensions there are
three kinds of modes, namely, extended, localized,
and multi-fractal separated by mobility edges.

However, the three-dimensional transition is different from the
bonded electron transition[1-10]. Specifically, there is no
divergence of the correlation length near the localization edge at $\omega_c$, 
and  the volume of the extended states (defined by the participation ratio)
converges to zero at the mobility edge. It is analogous
to the percolation phase transition above its threshold, where the 
global percolation cluster becomes a finite fraction of 
the size of the system.  Such a transition implies zero electronic
conductivity near the mobility edge,
which can also have a major effect on the heat conductivity of phonons
in glasses, where there are well-known anomalies
(for example, a plateau in heat conductivity in medium temperatures), see
for example Ref. \cite{FA}.

We have analyzed the eigenvectors of the two- and three-dimensional 
systems. Importantly, we found that in the two- and three-dimensional 
systems studied there are several scattering regimes. 
The first is at low frequencies in two and three dimensions
where there is no interaction between the modes, and they are characterized
by well-defined wave vectors.
The upper cutoff of this range depends on the system's scale.   
Above this cutoff the mapping between 
the state and the wave vector fails because of scattering. 
However, an absolute value of the wave vector is still defined for each state. 
The inverse scattering width is defined by the variance around 
this mean value.
This second range is called a `normal' scattering regime. 
We believe that it extends to zero
frequency when the system is enlarged.
Above this range beyond a point $\omega_b$ we find  anomalous scattering.
Above these three regimes there
are localized states where the wave length is hardly defined at all.
For normal scattering we find that the scattering width scales as
$\Gamma \sim \omega^{1.7\pm 0.1}$ in two dimensions and 
$\Gamma \sim \omega^{3.5\pm 0.5}$ in three dimensions
within a relatively small region of $\omega$.
%The third range is anomalous scattering range is very evident (strong).
In two dimensions there is an overlap between the appearance 
of anomalous scattering, the boson peak, and the localization edge i.e.
$\omega_c=\omega_b$. However,
in three dimensions there are two distinct scales.
The  first scale is the localization edge, and the second one is
located at the boson peak and at the anomalous scattering edge $\omega_b$,
where a drop-off in the participation is observed.
  
\section{ Localization and electrons}
We shall review briefly presented some background on localization and 
its relation to the electronic case. 
The classical example for such phenomena 
is an electron in a disordered potential. 
An electron that obeys 
the Schr\"odinger equation with spatial disorder in the voltage is known to 
be localized for a strong enough disorder. We examine the equation
\begin{equation}
i \dot \psi_j = \Delta \psi_j + V_j \psi_j,
\end{equation} 
where $j$ is the position, $\Delta$ is the Laplacian operator on the relevant 
lattice, and the potential $V_j$ is a random function. 
%\begin{equation}
%i\cdot \psi (r,t) = \grad^2 \psi(r,t) + V(r,t)
%\end{equation}
The variance  of $V$
 affects the character of the eigenstates,  the
conductivity, and its 
dependence on the system size[1-10].

In one dimension and for any type of disorder all electronic states 
are localized: the wave-functions decay at
large distances as 
\begin{equation}
<\psi(r)\psi(r+r_0)> \sim \exp(-|r-r_0|/\xi),
\end{equation}
where $\xi$ is the localization 
length, and $r$ and $r_0$ are any given lattice points. 
This leads to an exponential localization of the conductivity.
 
In two dimensions the situation is marginal and depends 
to a large extent on the models. For any disorder in $V_j$ the 
wave-functions form a multi-fractal set of the system.
The multi-fractal exponent is weakly dependent on the energy.
Above a certain critical disorder all electronic states are localized.

In three dimensions both extended and localized states exist.
There is a critical energy $E_c$ that defines the transition from localized to
extended states. 
This transition is accompanied by a diverging correlation length
$\xi(E) \sim (E-E_c)^{-\nu},$ where $\nu$ is a  positive number.  
In three dimensions, states are extended below or above a mobility edge and 
localized below or above it. The electronic 
conductivity is expressed as $\sigma\sim (E-E_c)^{\nu}$ \cite{thouless,KM93}. 
For electronic models with magnetic fields
there is also a localization transition in the 
two-dimensional case \cite{KM93}. 
Classical waves (photons and waves) were proposed to behave similarly
(see \cite{eco90,sheng89} and Ref. therein). 

There are various criteria for localization cited in the literature,
the first of which is the asymptotic behavior of localized states.
It is usually described in terms of the envelope of the 
wave-function as stated in Eq. 2. 

It is often helpful to consider the  moments of the wave-function values.
For a particular normalized  $\bar \psi_r(E)$,
a generalized participation ratio is defined as
\begin{equation}
I^{2q}(E) =  1/(\sum_r |\bar \psi_r|^{2q}(E)),
\end{equation} 
where the summation is done over all lattice sites, and $q$ 
is an arbitrary real number.
This measures the portion in space where the eigenvector 
is markedly different from zero. The dependence of $I^{2q}$ on $L$ can be
written as
\begin{equation}
I^{2q} \sim L^{D(q)},
\end{equation}
where $D(q)$ is a fractal exponent.
$I^{4}$ is the well-known participation ratio. 
The values of the $D$s  become constant in the limit of an infinite system.
If the states are extended, $D(2) = d$.
For  multi-fractal states  $I^{2q}$ is scaled
as $L^{D(q)}$ with $D(q)< (q-1)d$. For a localized state, $D(q>0) =0$ and 
$I^4\sim \xi^d$.  
 
Another useful measure of localization is the transport 
coefficients of the system.
In the electronic and thermal states the  transport is defined by the 
Kubo formulas or by transport definitions [8].
A localized electronic problem has zero conductivity at
zero temperature, whereas in multi-fractal cases it is believed that 
the conductivity converges to zero.
  
\section{ Models}
\subsection{Vibrations}

We begin this section with a discussion of the phonon model. 
If there is a particle
system that has a stable equilibrium, it is possible to displace it by
$\vec u_i $ from the stable particle positions. If the $\vec u$s are
small enough 
there is an elastic expansion around the 
equilibrium positions. 
Accordingly, vibration Hamiltonians of the following kind can be written as 
\begin{equation}
H = {1\over 2} \sum_{i,\alpha} m_i
(\dot u^{\alpha}_i)^2  + 
 {1\over 2}
%\end{equation}
%\begin{equation}
\sum_{i,j,\alpha,\beta}
%\end{equation}
%\begin{equation}
\Phi_{\alpha,\beta,i,j} u^{\alpha}_i u^{\beta}_j,
\end{equation}
where $u^{\alpha}_i$ is the $\alpha$ component of the translation in site $i$,
and $\Phi$ represents the interaction terms.

We shall limit our discussion here to scalar elastic movements
in an orthogonal direction to the state.  (This will inhibit the 
interaction between the transversal and longitudinal modes, which have
a role in vector models). 
So the discrete  elastic Hamiltonian can be given as
\begin{equation}
H = {1 \over 2} \sum_i m_i (\dot u_i)^2  +
  {1 \over  2} \sum_{ij} K_{ij} (u_i - u_j)^2,
\end{equation}
where $u_i$ represents  elastic movements and
$K_{ij}$ represents the elastic constants.

There are two obvious constraints on the parameters. 
The first is the symmetry $K_{ij} = K_{ji}$. The second is that  
$H$ must be non-negative for all realizations of $u$. 
This is a non-local demand for the elastic constants that is satisfied
by a positive $K_{ij}$.
The single atom dynamics is defined by
\begin{equation}
m_i \ddot u_i  = \sum_{j} K_{ij}(u_i-u_j),
\end{equation}
where $m_i$ is the particle mass.

Assuming that $u_i = \exp(i \omega t)\bar u_i(\omega)$
we obtain an eigenvalue problem represented by
\begin {equation}
-m_i \om^2 \bar u_i =  \sum_{j} K_{ij}(\bar u_i-\bar u_j).
\end{equation}
Note that the eigenvalue problem is generally not
 symmetric because of the existence
of the $m_i$ terms. 

\subsection{Diffusion and `Free' Quantum Particles}
Free random diffusion has a similar dynamical form.
We define particle densities $\rho_i$ and
random diffusivities $D_{ij}$ between the nearby sites. The 
 currents are $J_{ij} = D_{ij}(\rho_i-\rho_j)$ and we
obtain the random diffusion equation
\bel{model1}
\dot \rho_i = \sum_{j} D_{ij} (\rho_i-\rho_j). 
\ee
Defining 
$\rho_i(t,\mu) =\exp(-\mu t)\bar  \rho_i(\mu) $ results in
 a similar problem  with
$\mu = - \om^2.$ Since negative densities in $\rho$ are
not allowed, $D_{ij}$ should be greater than zero.
In contrast to the phonon model, there is no condition for detailed balance, 
and  $D$ does not have to be symmetric. 

A  free randomly moving quantum particle can be described by
\bel{model2}
i \dot \psi_i = \sum_{j} M_{ij} (\psi_i-\psi_j) 
\ee
for arbitrary Hermitian $M_{ij}$. For real and non-negative $M$s and for the   
linear solution $\psi = \exp(iE t) \bar \psi_i(E)$, 
we obtain a similar eigenvalue problem.
 
To summarize, the random realizations of
these three linear models have the same linear dynamical
structure, but only under the following conditions:
the transmission coefficients $M,D,$ and $K$
are symmetric, real, and positive, and  the phonon model
is scalar and possesses a single 
mass. All these models are Positive Random Laplacian models.
However, the time dependence, statistics
and other properties are different.
In this paper we use, 
unless stated otherwise, the phonon terminology.

The difference between these Random Laplacian models and
the electronic models is already manifested
in one dimension\cite{Ea,Derr84,ITZ10}. 
The single mass vibration model in one dimension can be mapped 
(see in the next section) to an electronic model with a disordered 
potential that decays with the square of
the frequency. In higher dimensions, however, we are not aware of 
any solution.
Since we know that this model is equivalent to the diffusion model, it 
is clear that for positive connection coefficients any density  will
 diffuse
to infinity. Therefore, the correlation length can
either explode at low frequencies or the system can be extended.

\section{Definition and simulation of the models}
\subsection{Definitions}
Here we focus on an equal mass ($m_i \equiv 1$) model in a
cubic lattice,  of $d$  dimensions, with a unit distance and  size $L$.
We considered nearest neighbor interactions, with two elastic 
constants $K_1$ and $K_2$, with
a ratio $\alpha = K_2/K_1$ between them, and a probability $p$ to 
find $K_1$ and $1-p$ for $K_2$. Since $\alpha<1$ is the same as 
$\alpha> 1,$ we will discuss only the former limit. 
The normalization $ K_1 = 1$ determines a scale of frequencies.
This model was suggested in \cite{CZ88}
and simulated in \cite{Ru91} for small $\alpha$s.  

There are two simple limits for this model, 
namely,  $\alpha =1$, which is  the ordered lattice limit, 
and $\alpha = 0$,  which is for percolation. 
In the ordered lattice limit, all states are extended and ordered.
The percolation limit, however, is more involved.
Above the percolation threshold, we normally find phonons on a scale
much larger than the percolation correlation
length \cite{ale82}, and  fractal excitations, named 
fractons are found below this length \cite{ale82}. 
Below the percolation threshold, the system would be completely
localized. 
For $\alpha > 0$, a different behavior exists.
With extremely small values of $\alpha,$ there are
fracton states together with phonons \cite{Ru91}. This  is 
not the case were the values of $\alpha$ are larger and
$p$ is not close to the percolation threshhold. 

We treated the eigenvalue problem numerically using 
the LAPACK package for real symmetric banded systems
for the equal mass model.
Specifically, we used a cubic system of size $L$, where the 
distance between particles is unity. 
Most of our simulations were performed under free boundary conditions.
However, we verified that the simulations were not changed under 
periodic boundary conditions. 
For this propose, we computed a set of eigenvalues $\om_i$s, 
and eigenvectors denoted as $\bar u_r(\om_i)$, 
where $r$ is the lattice position.
To verify the accuracy, we checked some examples for which the answers were
known and compared the results of other codes to our computation.

\subsection{The One-Dimensional Case}
The one-dimensional problem on an ordered lattice is 
a classic example  \cite{SS}.
Eigenstates can be written as
\begin{equation}
u_k = \exp(i(kr+\omega(k) t)).
\end{equation}
If the lattice is disordered the states can no longer be written
 as periodic
modes and thus the problem  becomes localized. 
This general problem was solved in
\cite{Ea,Derr84,ITZ10} and it is localized  with a 
localization length that is
proportional to $\omega^{-2}$. We thought that it would be useful to test our 
code on this system to check both the numerical code and the dependence 
of our results on the system size.
The solution of the theoretical model in one dimension can provide some 
insight as to why this model is so different from the bonded 
electronic model; therefore we present it in detail in Appendix A.

Considering the above, we simulated the one-dimensional problem.
Figure 1 displays our results for the participation ratio, which is
as noted previously, proportional to the correlation length.
This curve also shows an $\omega^{-2}$ dependence of the correlation length.
As the system size is enlarged, 
wider regions of such behavior can be observed in the curve. 
For a definite system size $L,$ there is a cutoff in the frequency
that is defined by $L \sim \omega_c^{-2}$. 
Below this cutoff the correlation length is larger 
than $L$. Since $\omega_c \sim L^{-0.5}$ and the
number of states is proportional to
$L$, there are $L^{0.5}$ free states in such a system.   
This can be observed in the plateaus on the right-hand side of the curve. 

\subsection{Two-Dimensional Models}
We now  analyze the previous model in two dimensions. 
As mentioned previously, for the electronic model
in two dimensions, any degree of variation in $V$ generates 
fractal exponents $D(2)$.
These effects are observed in system sizes between $16$ and $50$.
A transition in scaling for low noise levels can be easily observed.

In the two-dimensional positive Laplacian case we were
unable to distinguish any 
failure of normal scaling for small values of noise.
Therefore, unlike the electronic case, a small disorder has no 
effect whatsoever on the scaling described by the participation ratio.    
For smaller values of $\alpha$, localization is  indeed  observed.  
In Figure 2 presents numerical results of systems with parameters 
$p=0.4$ and $\alpha=0.1$. Figure 2a shows the density of states 
divided by $\omega$.
A boson peak with a center around $0.6-0.65$ is visible. 
Figure 2b presents $I^4$ as a function of $\omega$ for different 
system sizes. 
At low frequencies there are extended modes whose participation ratio
scales with the system size. There is a well-defined mobility gap 
at $\om_c$, separating the  extended and quasi-localized states.  
Below the mobility edge the participation ratio can be written as
$$I^4(\om) \sim |\om-\om_c| L^2.$$ 
All the states near the mobility edge are extended, though their
volumes may be very small.  
The density of states in the extended region is close
to the classical density $N(\om) \sim \om$. 
To analyze these states, we calculated  the exponent
$D(2)$. 
%
%In Figure 2c we show two scaling curves were we calculate 
%the scaling exponent $D(2)$ for two different frequencies.
% One observes a  scaling 
%over all the scaling range in $L$.  
%
Figure 2c shows the dependence of $D(2)$ on $\om$. 
Five scaling regimes of $D(2)$ are visible.
The multi-fractality of these states can be verified
either by calculating the dependences of $I^{2q}$ or
calculating of the distribution function of $\bar u_r$
and from point-point correlations.  
The transitions in $D(2)$ are sharp within our numerical accuracy.
Thus, there is a sharp  transition 
between the multi-fractal and extended states in this model at 
$\omega_c = 0.63\pm 0.03$. Note that this is also the  position
of the boson peak.

\subsection{Three-Dimensional Models}

The same kind of model was also simulated in three dimensions. Figure 3a
presents the density of states divided by $\omega^2$. A boson
peak is observed at $\omega \approx 0.6.$
Extended states and a localization transition to a fully localized state
were observed (see Fig. 3b).
Because of insufficient statistics, it is unclear whether there 
are quasi-localized states for this problem. 
Again, the scaling of the extended eigenstates near the localized 
regime behaves as
\begin{equation}
I^4 \sim |\om-\om_c| L^3.
\end{equation}
The states in the extended regime behave similarly to those e
in two dimensions.
Note that in this model the states up to a scale of 
$\omega\approx 0.65$ have a flat global scaling (see in Fig 3b). 
%%%%%%%%%%%%%%%%%%%%%%%%%%%%%%%%%%%%%%%%%%%%%%%%%%%%%%%%%%%%%%%%%%%%%%%

%%%%%%%%%%%%%%%%%%%%%%%%%%%%%%%%%%%%%%%%%%%%%%%%%%%%%%%%%%%%%%%%%%%%%%%

The phase transition from localized to delocalized states
is  different from that of the electrons. Specifically,
there is no divergence of the localization length, as is observed in the 
electronic models, so
$\xi \sim (\omega-\omega_c)^0$. This also indicates zero
`conductivity' below the transition
and will have an important effect on low heat conductivity
in glasses \cite{FA}.
The phonon volume in the extended regime shows a second 
order behavior similar to that observed for the volume
of the infinite cluster in percolation models.

Let us now compare in detail comparison of our numerical results with the 
theory. Note that both theoretical works \cite{SJ83,CZ88} are based
on approximations.  Here, comparison of a numerical simulation
with theory is possible, even for a system with a limited scale.  
The basic concept used in our simulations is the following: if there is
an increase in the system size, localization lengths that are 
not dependent on the system size will appear in the simulation
when the system scale becomes large enough.
This is indeed what was observed in our one-dimensional computation.
As the system size is increased, there is a decreased lower zone of frequency
where participation ratios depend on system size, and an increased 
range where the they fit the theoretical predictions.
This effect does not occur in two and three dimensions. There
is a clear localization transition in two dimensions, but the states above it,
even those with the smallest normalized participation ratios,
continue to scale with the system size. 
A two-dimensional  scaling length $\xi \sim \exp(-\omega^{-2})$ was 
predicted in \cite{SJ83}.
This indicates that above a cutoff given by $L\sim\xi,$ all states would scale
with the system size and there would be a shift in the cutoff when $L$
is increased. Both effects were not observed in our calculations.

\section{Scattering and eigenvector structure}
\subsection{The Two-Dimensional System}
As observed before the scattering of phonons and the shape of the scattering
is an extremely interesting issue.
We can expect to observe at least three
scaling ranges in the shape of the spectra.
Figure 4 presents shapes of various wave-functions 
(left) and the square of their Fourier transform 
${\bf F}(\vec k,\omega)$(right). 
Figure 4a shows for $\omega\approx 0.1$ a well-defined
periodic mode together with a well-defined peak 
in Fourier space. As the frequency is increased, 
there is an increase in scattering. 
This is manifested in the creation of local 
domains with different directions and
the existence of rings in Fourier space. 
At higher frequencies ($\omega\approx 2$),
the shapes become localized.   
Figure 5 gives examples of the radially integrated 
Fourier transform $F(|\vec k|,\omega)$ for the shapes presented in Figure 4.

To quantify these effects, we measured the
peak values of $k_0$, which  is scaled linearly with $\omega$ below the 
localization edge. Therefore, as stated before, they appear to be 
phononic. Above this edge over a limited region we find 
that $\omega \sim k^{0.85\pm 0.3}$.
Next, we assumed that the Fourier transforms can be represented as
\begin{equation} \label{fit}
F(k,\omega) = \frac {\Gamma^2} { (k-k_0)^2 + \Gamma^2}.
\end{equation}
We make a fit with this expression to try to determine the 
amount of scattering. 

Fits with this expression are reasonably good up to a value 
of $\omega \approx 1$. 
Figure 6a shows the linear dependence of $k_0$ at a
frequency up to  $\omega_b \approx 0.65$. Above this range the Lorentzian fits
are less good because of the existence of larger tails at large $k$s.
In this range we observe (see Fig. 6a) a 
transition to an anomalous dependence of the wave vector
 $\omega \sim k^{0.8 \pm 0.05}$. 
Figure 6b shows the width $\Gamma$ as a function of $\omega$.
In the smaller range scale,
for small frequencies up to a scale of 
$\omega = 0.2\pm 0.05,$ the width is constant, since 
the system is not large enough to create scattering at these wavelengths.
Above this value up to $\omega_b= 0.65,$ we obtain a dependence 
of $\Gamma \sim \omega^\mu$
with $\mu,$ where $\mu = 1.7 \pm 0.2$. In the second frequency 
range there is a smaller exponent, $\mu=1.1 \pm 0.1 $.  

This transition approximately
overlaps the change between extended and multi-fractal modes
: $\omega_b=\omega_c$.
For the higher frequencies, the states are more
localized and fits to  Eq. \ref{fit} are not good.

\subsection{The Three-Dimensional Case}
We performed the same type of fit in this case as in the two-dimensional
case. The results are presented in Figure 7. Here the range
in Fourier space is smaller (one decade) and this makes some of our
estimates inaccurate.
As in the two-dimensional case, we observe for small $\omega$ values
a range of constant peak width. 
For larger values of $\omega$ there is a range with a very sharp rise, 
which can be fitted to $\Gamma \sim (\omega-\omega_c)^{3.5\pm 0.5}$. For
$\omega_b \sim 0.66$ there is a crossover to anomalous scattering.
When $\omega$ is further increased, there is a transition
to localized states. We noted that this 
usually occurs in glassy models where the transition
to a strong scattering transition happens within the 
extended range.

Note that the dependence on $\Gamma$ that we would obtain
in optical Rayleigh scattering is $\Gamma \sim \omega^{d+1}$, 
though this is not the same type of problem for the 
phonons\cite{Courtens}.
Our results in two dimensions are not consistent with this.
In three dimensions they might be  consistent with such scattering.

In two dimensions the transition between normal 
and anomalous scattering is exactly the point where
the localization transition and the boson peak is observed.
All these effects are clumped together in two dimensions. 
%However, in three dimensions we have observed that the
%localization point is above
%the position of the  anomalous scattering and boson 
%peak. 
In three dimensions there are two transitions: the first is the
localization edge $\omega_c$ and the second is at the transition between normal and 
anomalous scattering and  the boson peak at $\omega_b$. 
This is a strong indication that in three dimensions,
the strong scattering effects begin before the localization edge effects do.
This is also in accord with the commonly accepted situation as indicated
by the literature on glasses. 
%Strong scattering starts to occur
%below the localization edge. 

\subsection{Relation to Experiments and General Discussions on Glasses} 

In actual experiments the dynamical correlations are measured by
\begin{equation}
S (\vec k,\omega) = 
\int_{-\infty}^{\infty} dt \exp(-i \omega t) <\sum_{i,j} 
\exp(i \vec k\cdot (r_i(t)-r_j(0)))>,
\end{equation}
where the average is done in time at
the relevant measurement temperatures.  
If we have a complete knowledge of the eigenvectors we can transform this
expression to
\begin{eqnarray}
S(\vec k,\omega) = & {1\over{\omega^2}} & \sum_l  \delta(\omega- \omega_l) 
N(\omega,T)  \nonumber \\ 
              &                    & \sum_{ij} e^l_i\cdot k e^l_j\cdot k 
\exp(i k (\vec r^0_i-\vec r^0_j))\exp(-W_{ij}) \nonumber \\
              &                    &  + h.o.t,
\end{eqnarray}
 where the $e$s are eigenvectors, 
$N(\omega,T)$ is the distribution function and the $W$s are 
the Debye-Waller factors. The basic components in this sum
are
\begin{equation}
{\bf G}(\vec k,\omega_i) =  k^2 {\bf{F}}  ( \vec k, \omega_i).
\end{equation}
Usually the sum is done over some range in $\omega,$ so it represents 
some average over the modes. In the normal scaling region the results 
we get from both kinds of fits are in accord. 
In two dimensions the fits can
fail because of the existence of a longer tail in the Fourier transforms.

In this paper we showed that indeed as speculated in the literature there
in 3 dimensions there is a cross over frequency between anomalously
 scattered waves 
and weak scattering in which the width of the square of the fourier 
scales as $\omega^{3.5\pm 0.5}$. The scale of this crossover is not the
Ioffe-Rogel scale so we don't know if it makes any sense to use this
terminology. The conclusions from
such a theory to transport is not clear. 
The reader can read some thoughts about it and some refrences
from the glass literature in \cite{Courtens22}. The transprt situation for 
states below the cross over should be discussed using different theoretical
measures then ours. Nonlinear effects cannot be approximated using our method.

We note that if there are nonlinear scattering 
effects associated
with vibrations at low temperatures (like two-level systems),
they will be observed in scattering experiments but not in harmonic analysis.
Furthermore if there are lifetime effects they will not be observed in such an
experiment. 

As noted before, there is no actual available data for the normal 
scaling range, neither in real glasses nor in the numerical simulations, 
because of various experimental and numerical difficulties,
so it is impossible to compare our results with experimental data. 
We hope that in due time this 
will be possible.

\section{Conclusions}

In this paper we focused on 
the relatively simple case of the scalar vibration model.
This can be demonstrated experimentally, for example, on a membrane. 
There are nontrivial predictions on the localization in such a system.
We observed that the eigenvectors for these models are strongly scattered
and that there are strong scattering regions in the extended regime
when the problem is in two and three dimensions. 
The dependences of the  width of the Fourier transforms of the
modes do not seem to be in accord with naive Rayleigh scattering estimates
in two dimensions.

We find that in the two-dimensional system there is only one
critical scale that is observed  by a localization edge, 
a boson peak, and a shift in the scattering and the 
dependence of $\omega$ on $k$. 
On the other hand, in the three dimensional system there are at least
two scattering scales and three scattering regimes.
The first one is the localization edge and the 
second is the scattering
edge where scattering becomes large.
The second scale overlaps with the boson peak.
A similar picture about Silicon glass 
is given in \cite{Pa99} even though we do not agree
with many of the statements in this paper about the nature of the states. 
This subject will be discussed in more detail in 
another paper where the transport coefficients will be calculated.
   
Introducing of vector disorder and structural disorder will induce
further complications, consisting of interactions between longitudinal 
and transversal modes. 
The existence of  mass disorder is another strong mechanism for 
introducing more scattering, especially in the optical bands
This subject is especially interesting because the representation
of systems with mass disorder is equivalent to a Schr\"odinger equation
with a correlated potential. This will be discussed in another paper\cite{OG99}. 
Another subject that remains largely untreated in our paper
is the definition of electrical 
and thermal conductivities. Kubo and transport formulas can be used to estimate
the dependence of these conductivities on temperature and elastic disorder
\cite{Kubo}, 
this will be done in another study. 

We demonstrated that a set of models  
has a different behavior with the effect of noise than bonded electrons.
This is manifested in two main ways: (1) the phonon 
models seem to be delocalized in two dimensions, and  
(2) they have a percolative behavior as a function
of the frequency. The scale of the mode diverges as the distance from the 
critical points times the system volume. 

We focused on the relatively simple case where the movements are scalar.
This can be realized experimentally, for example, on a membrane. 
There are nontrivial predictions for such a system.
Introducing of vector disorder and structural disorder will induce
further complications, consisting of interactions between longitudinal 
and transversal modes and possibly soft modes (to be discussed elsewhere). 
The existence of  mass disorder is another strong mechanism for 
introducing more scattering, especially in the optical bands. 
Notably, the eigenvectors for these models are strongly scattered and 
there are strong scattering regions in the extended region in two and three
dimensions. 

Another subject that remains largely untreated in this paper
is the definition of electrical 
and thermal conductivities. Kubo formulas can be used to estimate
the dependence of these conductivities on temperature and
elastic disorder\cite{Kubo},
which will be done in another study.
   
We acknowledge discussions with O. Gat, J.P.Eckmann, S.R. Elliott
E. Courtens and R. Vacher. 
???
%T. Kustanovich,        
%R. Zeitak O. Gat 
for discussions of the subject. 
\pagebreak

\bigskip
\appendix
\section{ The One-Dimensional Vibration Problem}

It was shown in  \cite{Derr84,ITZ10} that the phonon model is equivalent 
to the electronic model with a rescaled interaction
$V_i \omega^{-2}$. We shall briefly repeat the arguments
because, first, we can use them to solve
this problem and second, because it is reasonable to assume that
there is an analogous effect at higher dimensions. 

The one-dimensional equation can be written as

\begin {equation}
\om^2 \bar u_i =  K_{i}(\bar u_i-\bar u_{i+1})+K_{i-1}(\bar u_{i-1}-\bar u_i).
\end{equation}
This model can be solved through the following transformations:
if we define $Q_{n} = K_{i} (\bar u_i-\bar u_{i+1}),$
the equation is transformed to
\begin{equation}
Q_{n+1}-2Q_{n} +Q_{n+1} = -\Omega /K_n Q_n,
\end{equation}
where $\Omega = -\omega^2$.
This is a Schr\"odinger equation with a potential that scales down with
the frequency.  
Eigenvalues explode exponentially as
\begin{equation}
Q_n = \exp (n \gamma(\Omega)).
\end{equation}
An equivalent equation for $R_n = Q_{n+1}/Q_{n}$ can be written as
\begin{equation}\label {map}
R_n = 2+ \Omega /K_i -1/R_{n-1}.
\end{equation}
We can now write an equation for the probability distribution
$P(R)$ in the steady state. 
Here we get
\begin{eqnarray}
P(R) = & {p \over 2+ \Omega /K_1 -1/R} P( 2+ \Omega /K_1 -1/R) &+\nonumber \\
 & { 1-p \over 2+ \Omega /K_2 -1/R} P( 2+ \Omega /K_2 -1/R) &.
\end{eqnarray}
The localization exponents are defined as 
\begin{equation}
\gamma(\Omega) = \int P(R) ln(R) dR.
\end{equation}

For small values of  couplings or frequency 
the solution is defined by a fixed-point  solution. Using the 
average value of the map $b = \Omega \bar A$, 
($\bar A = (p/K_1+(1-p)/K_2)$) we get 
\begin{equation}\label {map}
\bar R = 1+ \frac {b}{2} +\left( b+\frac{1}{4}b^2\right)^{1\over 2},
\end{equation}
which can be expanded to 
\begin{equation}
\bar R-1 = 1+ b^{1 \over 2} + {b\over {2}} + O(b^{3/2}).
\end{equation}
The localization length follows 
\begin{equation}\label{1loc}
\xi = 8/b ={8 \over \bar A } \omega^{-2}.
\end{equation}
 
These results can be tested in a simulation as we here, mainly,
to test the simulation.
 
%%%%%%%%%%%%%%%%%%%%%%%%%%%

\onecolumn
\begin{figure}
\epsfxsize=18.truecm
\epsfbox{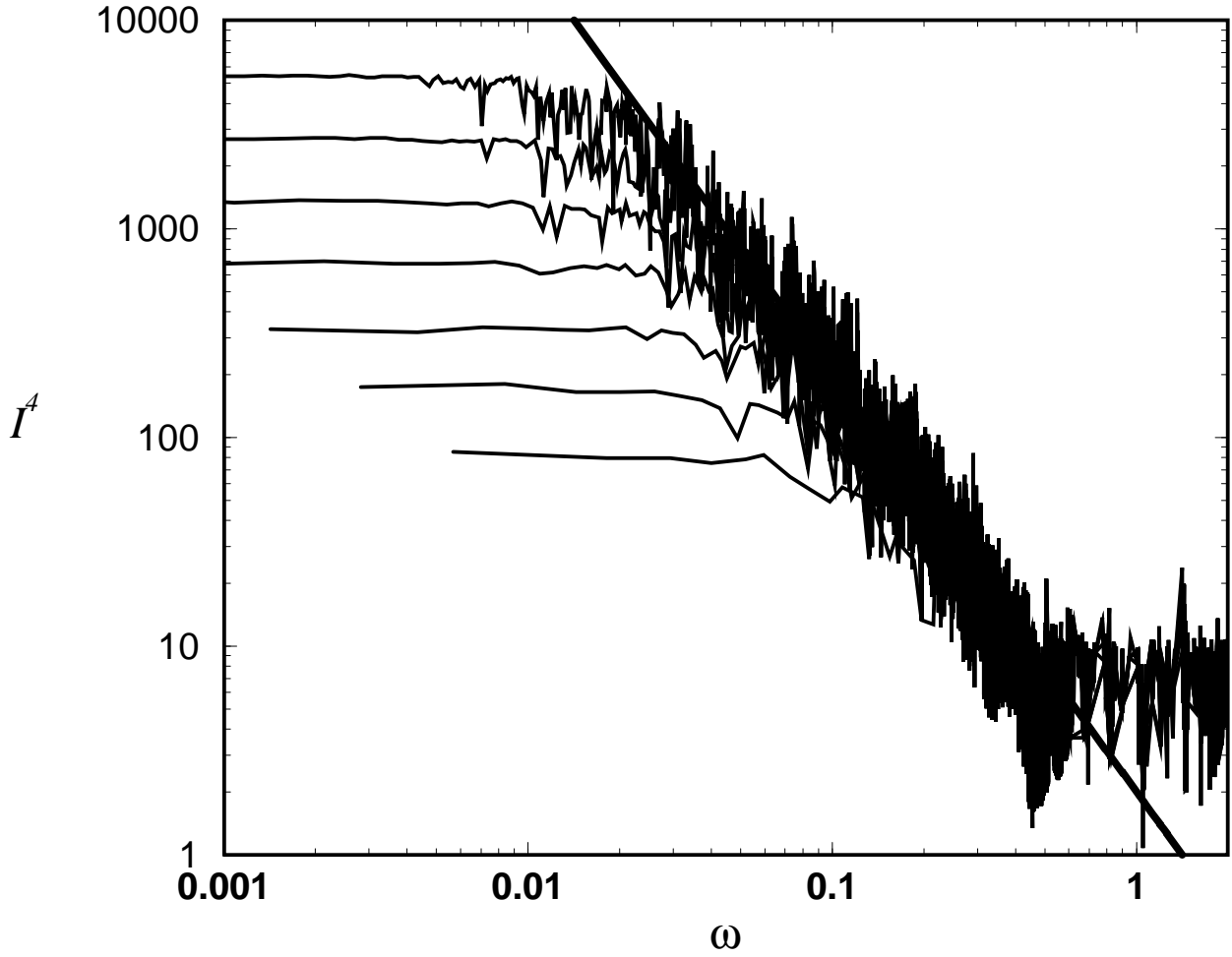}
\vskip 0.2cm
\caption{
Log-log scale of the participation ratio $I^4$ (proportional to $\xi$) 
as a function of system size 
for one-dimensional systems with $\alpha = 0.1$ and $p = 0.4$,
for sizes 128-8192 with jumps of factor two. We observed that the curves decay
as $\omega^{-2},$ as the bold line illustrates. 
As the system size is increased, this scaling
range also increases. Note that the upper cutoffs are defined by
$L \sim  \omega_c^{-2}$. Below $\omega_c,$ $I^4$
is proportional to the system size.
}
\label{Fig1}
\end{figure}
%%%%%%%%%%%%%%%%%%%%%%%%%%%%%%%%%%%%

\begin{figure}
\epsfxsize=8.truecm
\epsfbox{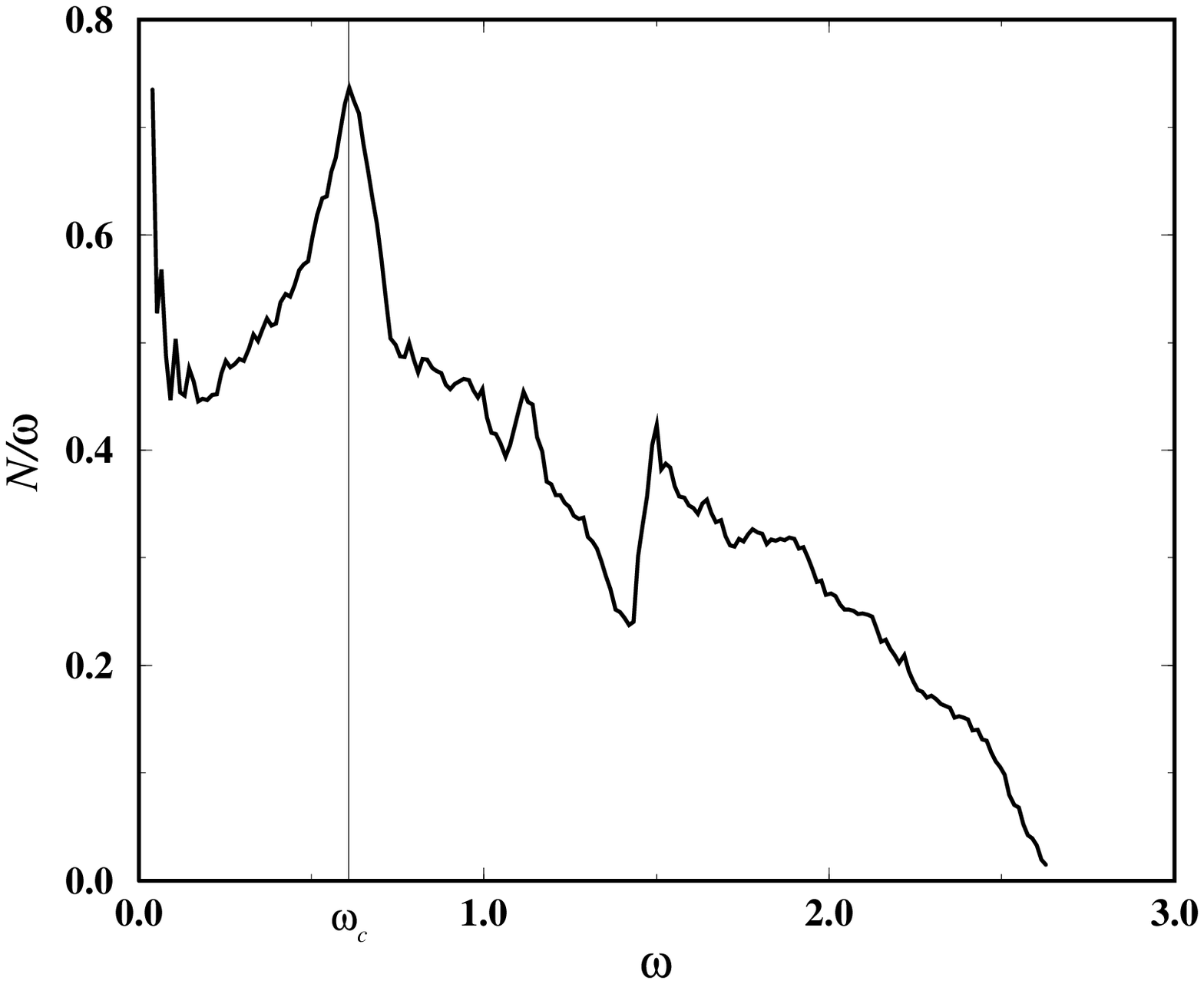}
\epsfxsize=8.truecm
\epsfbox{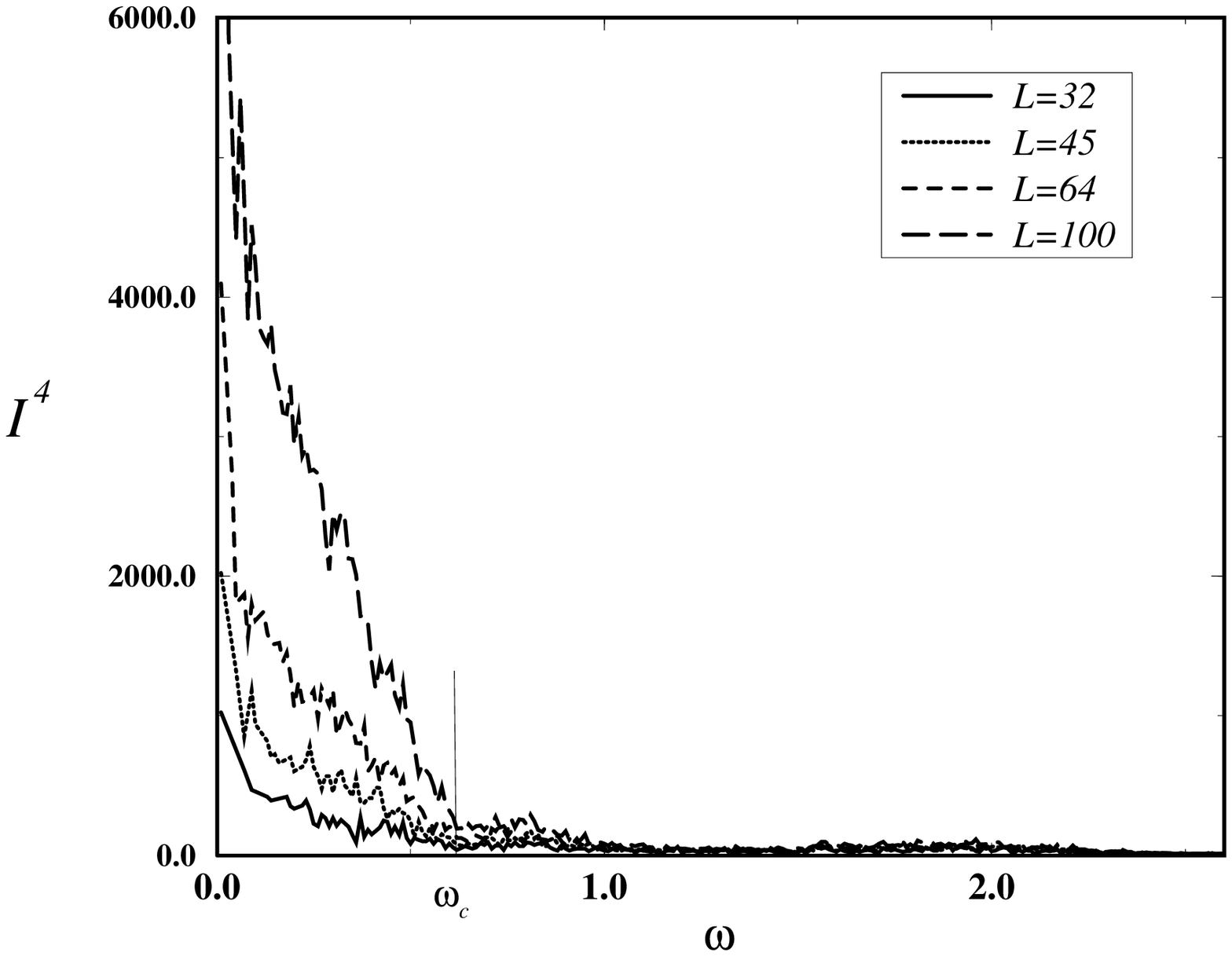}
\epsfxsize=8.truecm
\epsfbox{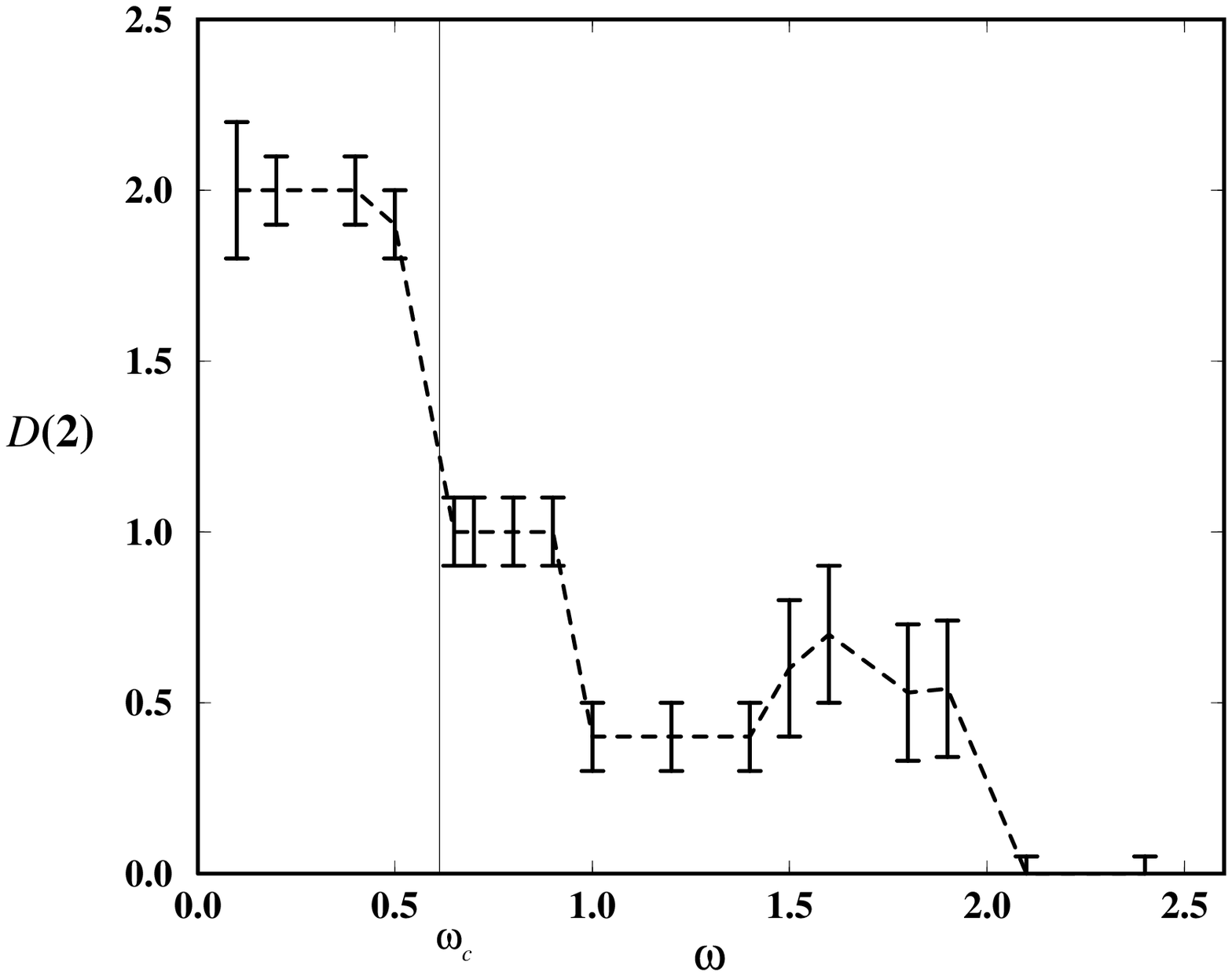}
\vskip 0.2cm
\caption{
(a) The density of states $N$, for $p=0.4$ and $\alpha=0.1$, divided 
by $\omega$ for a system size $L=100$.
The peak is the extra number of states relative to the linear normal
distribution, which is called the boson peak.
(b) The participation ratio is $I^4$.
The curves are averaged over equal intervals of $\omega$ of size $0.05$
to reduce  the $I^4$ fluctuations. 
The extended modes are seen  below $\omega \sim 0.63$. 
(c) The scaling exponent $D(2)$ is shown as a function of the 
frequency $\omega$.
The curve is averaged over intervals of frequencies $\omega$ in order to
reduce the noise in  $I^4$.
}
\label{Fig2}
\end{figure}

%%%%%%%%%%%%%%%%%%%%%%%%%%%%%%%%%%%%%%%%%%%%%%%%%%%%

\begin{figure}
\epsfxsize=8.truecm
\epsfbox{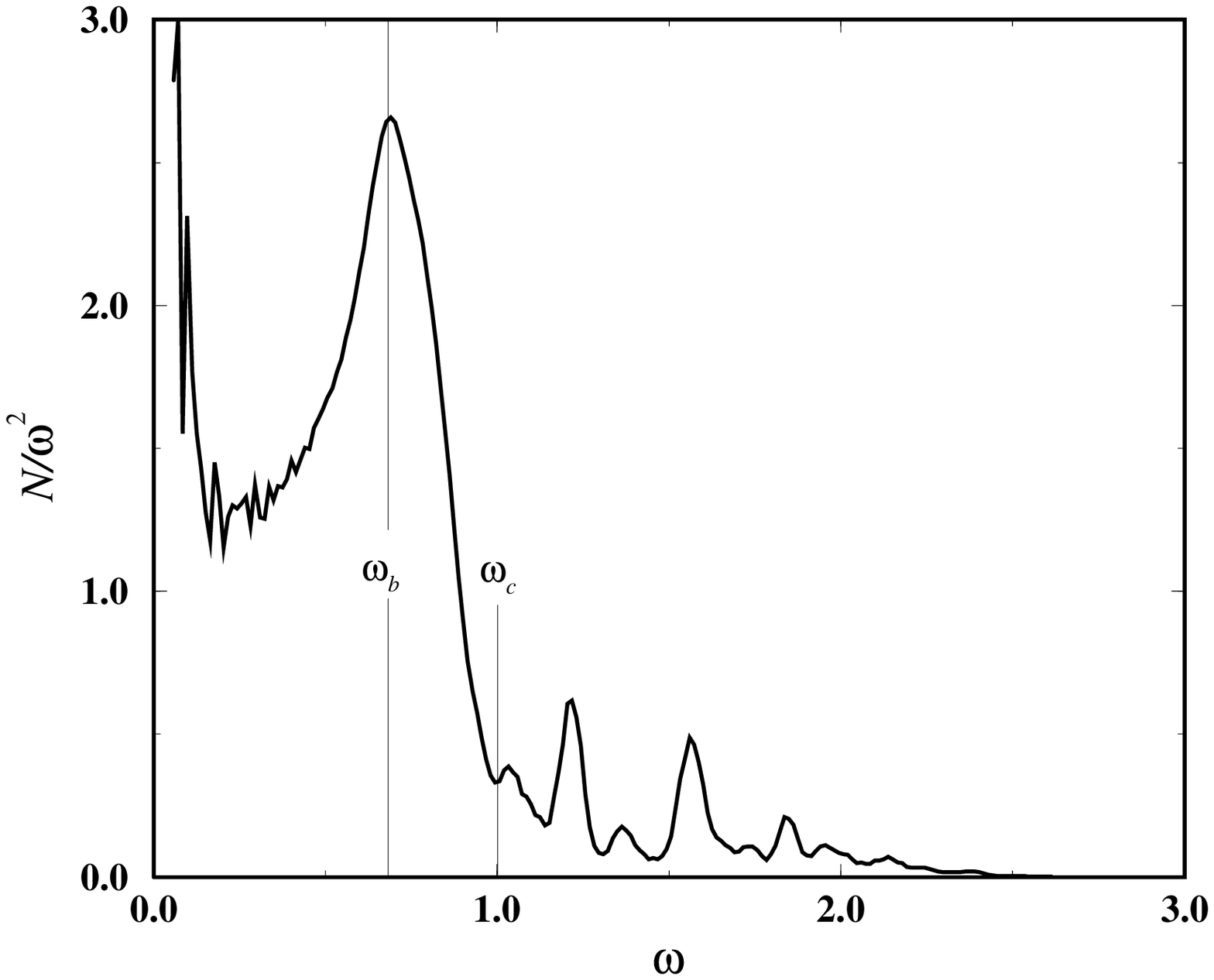}
\epsfxsize=8.truecm
\epsfbox{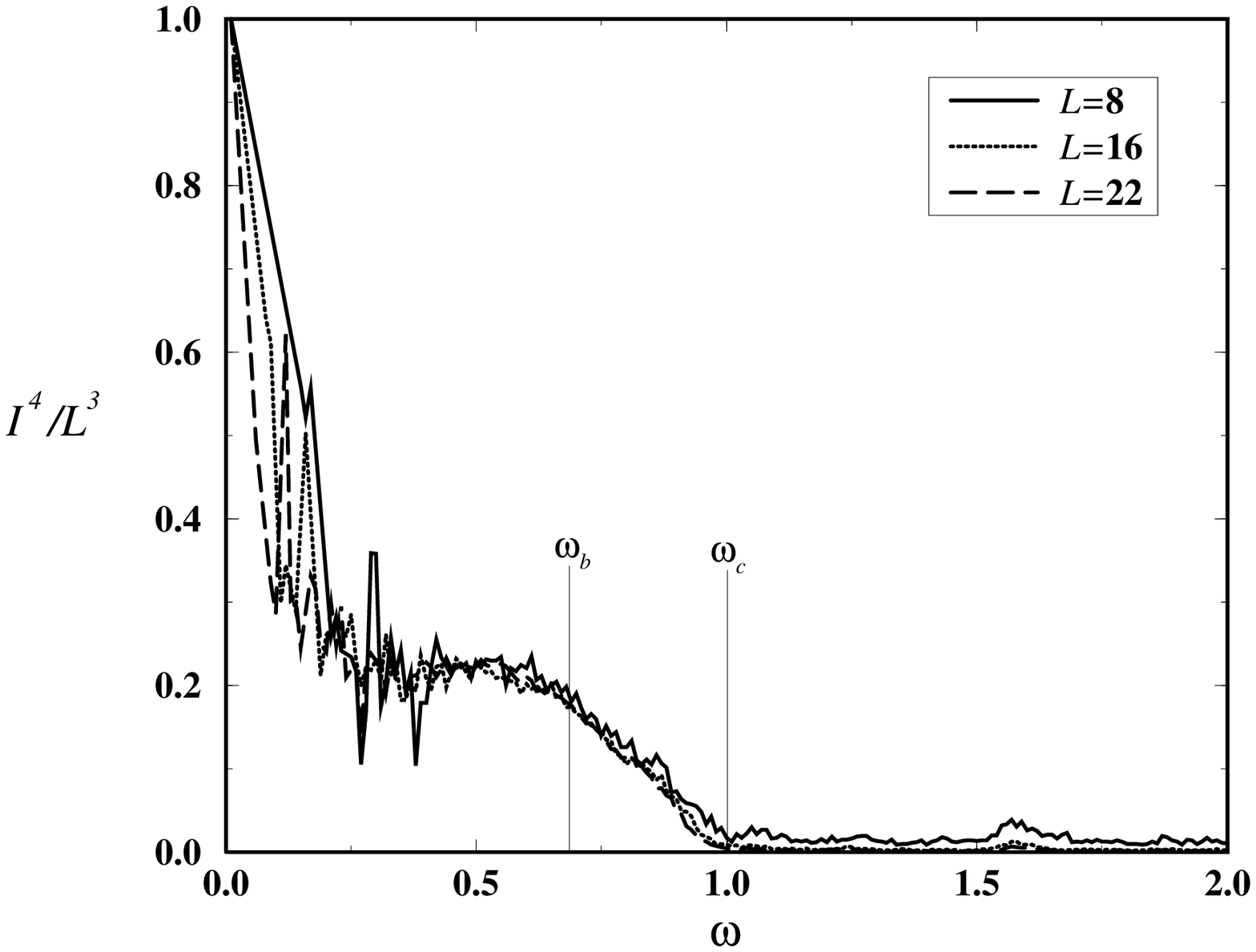}
\vskip 0.2cm
\caption{
(a) The density of states $N$, for $p=0.8$ and $\alpha=0.1$, divided 
by $\omega^2$ for a three-dimensional system with a size of $L=22$.
The peak is the extra number of states relative to the linear normal
distribution, which is called the boson peak.
(b) The participation ratio $I^4$ normalized by $L^3$ with the same parameters.
The solid line is for $L=8$, dotted for $L=16,$ and dashed for $L=22$.
The curves are averaged  to reduce the  fluctuations. For the system
with $L=8,$ we ensemble the average of 6 runs to reduce fluctuations.
The extended modes are observed below $\omega \sim 1$ by their collapse 
into one curve. 
}
\label{Fig3}
\end{figure}

%%%%%%%%%%%%%%%%%%%%%%%%%%%%%%%%%%%%%%%%%%%%%%%%%%%%

\begin{figure}
\epsfxsize=18.truecm
\epsfbox{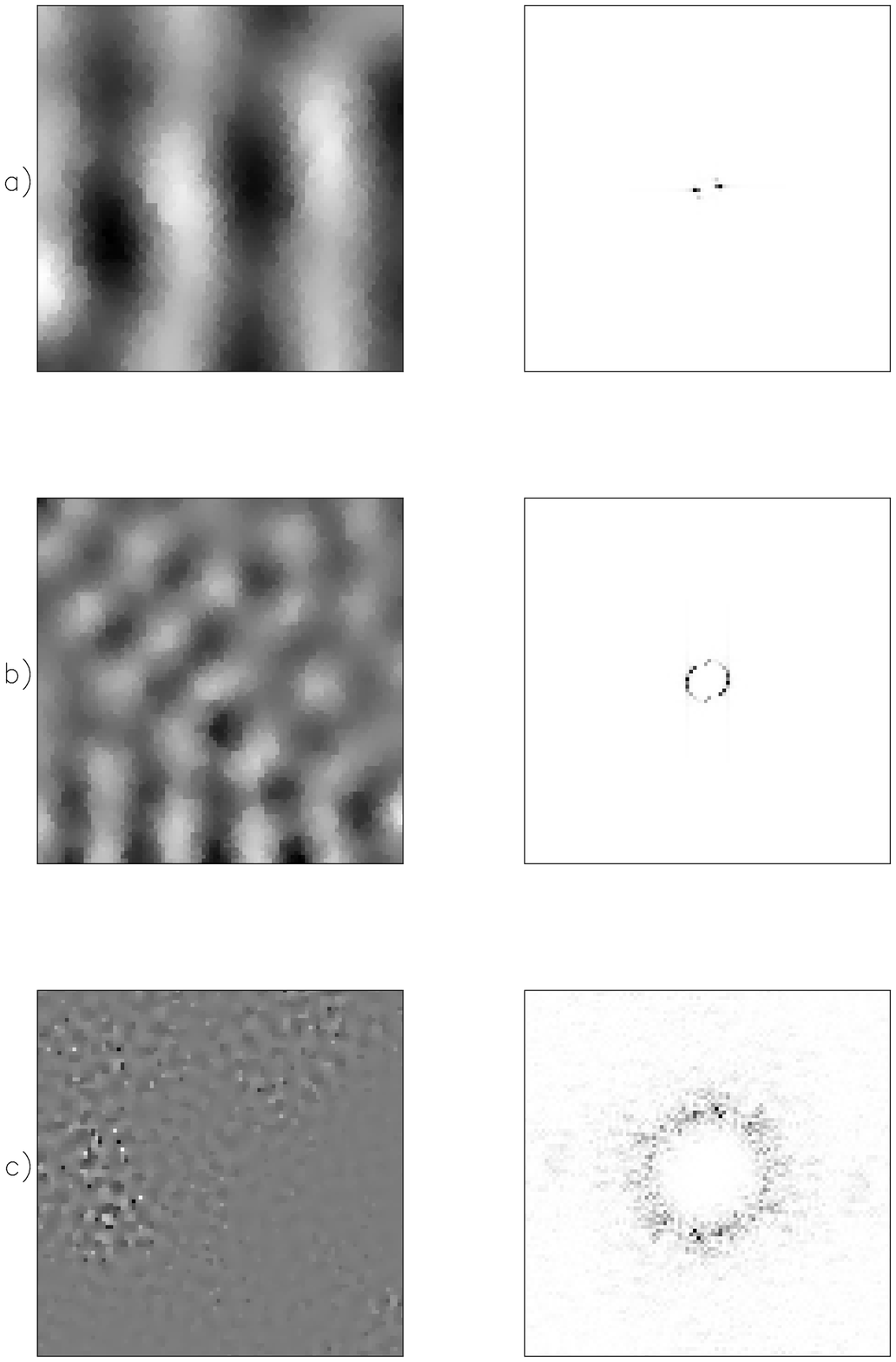}
\vskip 0.2cm
\caption{ We present the two-dimensional shapes and Fourier transforms of
the states of the $100^2$ system with $p=0.4$ and $\alpha=0.1$ 
for $\omega = 0.1(a),0.2(b), 0.6(c),1.0(d),2.0(e)$.  
}
\label{Fig4}
\end{figure}

%%%%%%%%%%%%%%%%%%%%%%%%%%%%%%%%%%%%%%%%%%%%%%%%%%%%%%%%%%%%%%%%%%%%%%%

\begin{figure}
\epsfxsize=18.truecm
\epsfbox{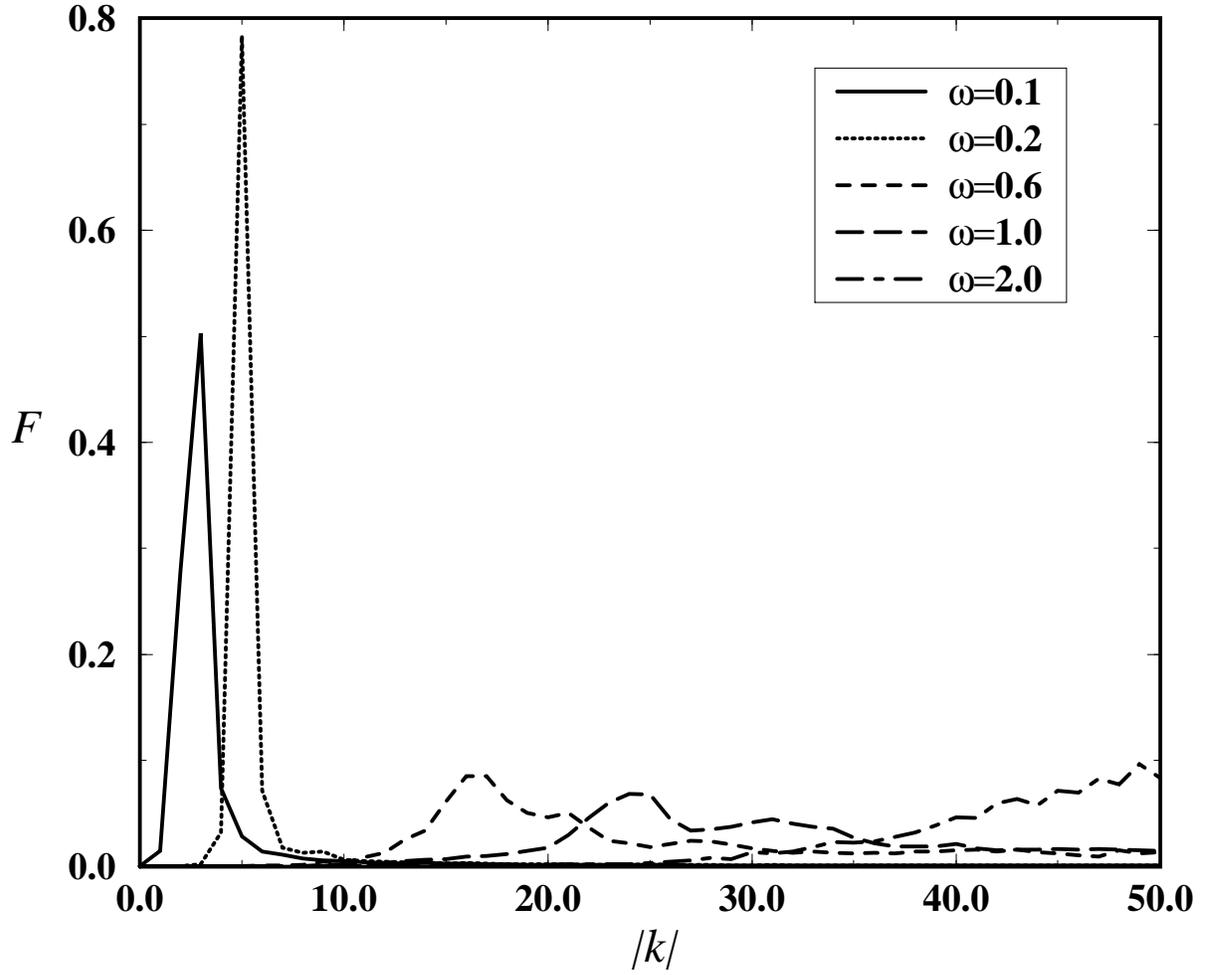}
\vskip 0.2cm
\caption{The radially averaged squares of the two-dimensional 
Fourier transforms $F(k)$ for the previous $\omega$s are shown.
}
\label{Fig5}
\end{figure}

%%%%%%%%%%%%%%%%%%%%%%%%%%%%%%%%%%%%%%%%%%%%%%%%%%%%%%%%%%%%%%%%%%%%%%%
\begin{figure}
\epsfxsize=18.truecm
\epsfbox{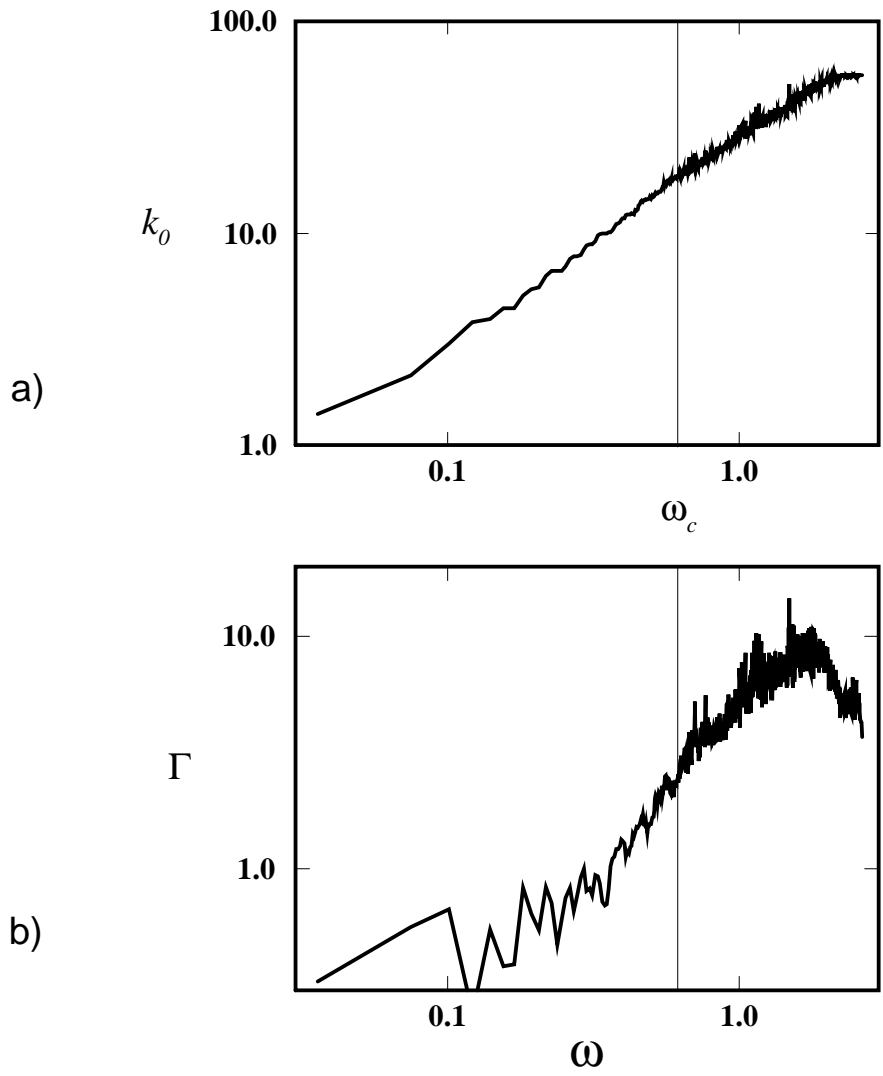}
\vskip 0.2cm
\caption{ The length  $k_0$ of the Fourier transforms as a function of
$\omega$ (a) and the width
$\Gamma$ of the peak (b) for the same two-dimensional system.
}
\label{Fig6}
\end{figure}

%%%%%%%%%%%%%%%%%%%%%%%%%%%%%%%%%%%%%%%%%%%%%%%%%%%%%%%%%%%%%%%%%%%%%%%

\begin{figure}
\epsfxsize=18.truecm
\epsfbox{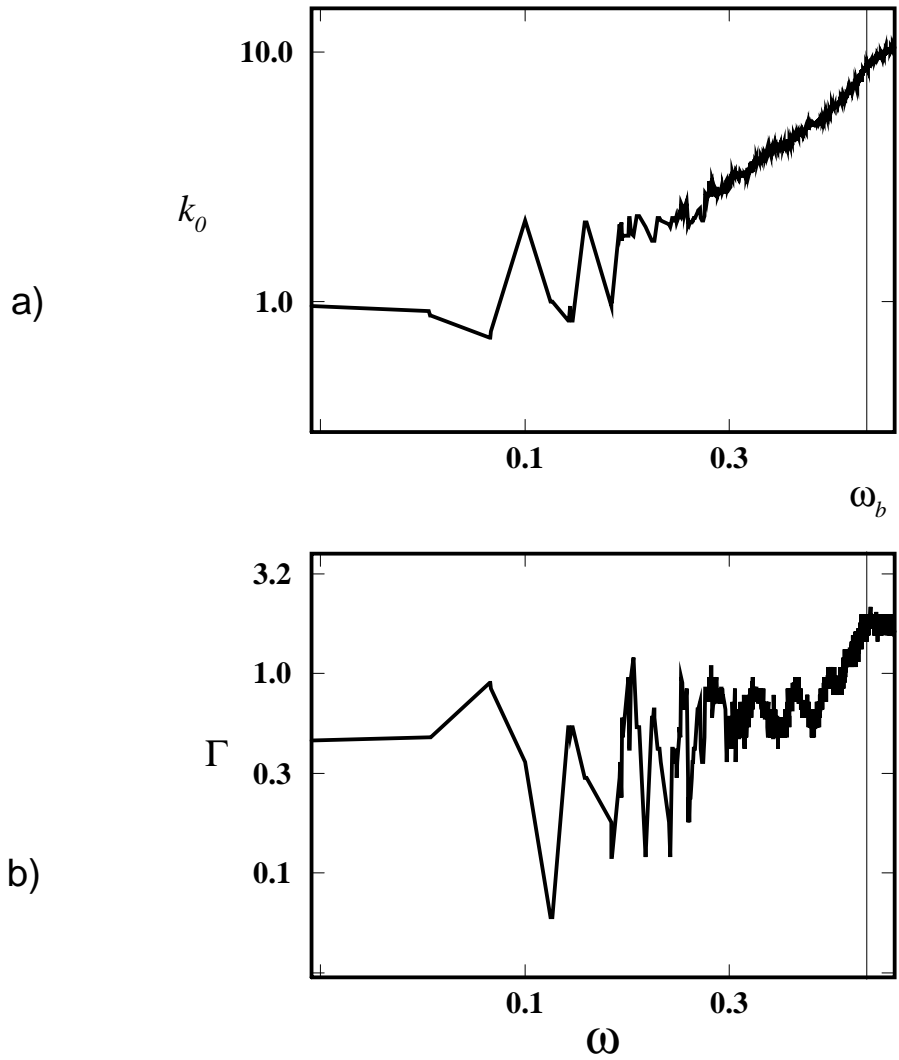}
\vskip 0.2cm
\caption{ The length  $k_0$ of the Fourier transforms as a function of
$\omega$ (a) and the width
$\Gamma$ of the peak (b) for the three-dimensional system with $p=0.8$,
$\alpha=0.1,$ and $L=22$.
}
\label{Fig7}
\end{figure}

%%%%%%%%%%%%%%%%%%%%%%%%%%%%%%%%%%%%%%%%%%%%%%%%%%%%%%%%%%%%%%%%%%%%%%%

\end{document}